\documentclass[10pt]{iopart}
\usepackage{subcaption}
\usepackage{graphicx}
\usepackage{citesort}
\usepackage{xcolor}

\usepackage{iopams}  
\usepackage{tabularx}
\newcolumntype{Y}{>{\centering\arraybackslash}X}
\newcommand{\phyp}{\textit{Physarum polycephalum~}}
\renewcommand{\etal}{\textit{et al~}}

\newcommand{\rev}[1]{#1}
\newcommand{\text}[1]{\mathrm{#1}}

\graphicspath{{./images/Images_Raf/}}

\begin{document}

\title{Emergence of dynamic contractile patterns in slime mold confined in a ring geometry}

\author{Valentin Busson$^{\ddag}$, Raphaël Saiseau$^{\ddag}$ \& Marc Durand}
\footnote{These authors contributed equally to this work.}


\address{Laboratoire MSC, Universit\'{e} Paris Cité, CNRS, UMR 7057, Mati\`{e}re et Syst\`{e}mes Complexes (MSC), F-75006 Paris, France}
\ead{marc.durand@univ-paris-diderot.fr}
\vspace{10pt}
\begin{indented}
\item[]March 2022
\end{indented}

\begin{abstract}
Coordination of cytoplasmic flows on large scales in space and time are at the root of many cellular processes, including growth, migration or division. These flows are driven by organized contractions of the actomyosin cortex. In order to elucidate the basic mechanisms at work in the self-organization of contractile activity, we investigate the dynamic patterns of cortex contraction in true slime mold \phyp confined in ring-shaped chambers of controlled geometrical dimensions. 
We make an exhaustive inventory of the different stable contractile patterns in the absence of migration and growth. We show that the primary frequency of the oscillations is independent of the ring perimeter, while the wavelength scales linearly with it. We discuss the consistence of these results with the existing models, shedding light on the possible feedback mechanisms leading to coordinated contractile activity.
\end{abstract}

%
\vspace{2pc}
%
\submitto{\JPD}
%
\maketitle
%
\ioptwocol

\section{Introduction}

Many biological processes such as cell migration, cell
division, or development are related with a transport of cytoplasm over long distances. Such coordinated flows cannot be produced by simple passive diffusion, but must be driven by active stress generation from cytoskeleton \cite{alim_mechanism_2017}. Locomotion of many organisms, such as slime molds \cite{slimes_book} and amoebae, relies on the generation of contraction waves along their body, thereby inducing cytoplasmic flow and hence body mass transfer. In absence of any external cues, such a mechanism gives rise to locomotion with anomalous diffusion behavior (persistent random walk) \cite{Rodiek_2015,Cherstvy_2018}. 
However, it is unclear how cortex dynamics can self-organize to generate to coordinated flows across large scales in space and time. The relative importance of chemical and mechanical  factors remains largely unknown. 

%
Particularly, \emph{Physarum~polycephalum}, a true slime mold, is renowned for long-range coordination of actomyosin cortex contractions in its plasmodial stage, driving giant cytoplasmic flow through a dense tubular network that can be seen with naked eye.
It is then widely used as a model organism for studying flow-driven amoeboid locomotion as well as the dynamics of self-organized cytoskeleton contractions \cite{rieu_periodic_2015}.
The plasmodium consists in a giant single cell with multiple nuclei and is composed of a highly invaginated outer layer of gel protoplasm (the ectoplasm) and a sol-like inner protoplasm (the endoplasm). Its giant size and two-dimensional growth facilitate the observation of thickness variation and cytoplasmic flows. The many complex thickness patterns visible to the naked eye that spontaneously emerge have motivated the development of numerous theoretical models. Although they all agree on the role of calcium as a regulator of the cytoskeleton contractions \cite{oster_mechanics_1984,teplov_continuum_nodate,oettmeier_physarum_2017}, they have a diverging opinion about the mechanistic
details connecting the calcium oscillations with the cytoskeleton
dynamics. For some, the contractility and the
resulting protoplasmic flows need to be considered as a part of the
feedback loop leading to calcium oscillations  \cite{kobayashi_mathematical_2006,radszuweit_model_2010,radszuweit_active_2014}. For others, bio-chemical reactions lead an autonomous calcium oscillator prior to any cytoskeleton contractions of flow \cite{teplov_continuum_nodate,teplov_role_2017,radszuweit_intracellular_2013,julien_oscillatory_2018}. \rev{For simplicity, we will refer to the first kind of models as \emph{mechano-chemical}, and the second as   \emph{mechano-bio-chemical}.}

On the experimental side, quantitative descriptions of the stable contractile patterns are scarce, and limited to homogeneous protoplasmic droplets \cite{TAKAGI2008420,TAKAGI2010873} or fusing plasmodia \cite{tero_coupled-oscillator_2005}, and hence can hardly discriminate between proposed models.



In this paper, we investigate the dynamic contractile activity of \emph{Physarum polycephalum} plasmodium confined in annular-shaped chambers of controlled geometrical dimensions, 
allowing to reduce the problem to a quasi-unidimensional system, but preserving the structural heterogeneities of a (macro)plasmodium. Moreover, the periodic boundary conditions suppress antero-posterior axis and so pre-established polarity of the giant cell, while confinement and deprivation of nutrients hinder plasmodium growth or displacement.
%
We take advantage of these simplifications to make an exhaustive inventory of the different stable contractile patterns that emerge spontaneously, and study the effect of geometrical confinement on the pattern properties.  We compare these observations with the predictions of existing theoretical models \cite{radszuweit_model_2010,radszuweit_intracellular_2013,radszuweit_active_2014,teplov_role_2017,julien_oscillatory_2018}. We show that our results are in favour of a mechano-bio-chemical coupling in which self-sustained calcium oscillations are prior to the emergence of spatio-temporal coordination of the contractions.

\begin{figure}[h]
	\centering
	\includegraphics[height=\columnwidth]{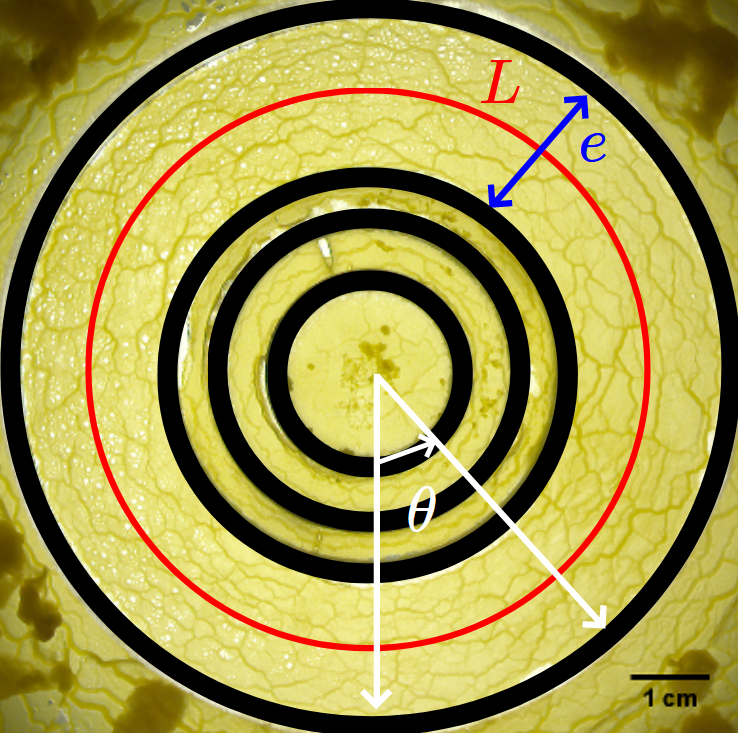}
	\caption{Raw image of an initially homogeneous \phyp plasmodium cropped with four concentric annular punches (in black). The ring perimeter $L$, width $e$ and angular position $\theta$ are indicated on the Figure.}
	\label{fig:Set-up_image}
\end{figure}

\section{Results}

\subsection{1D description of ring-shaped plasmodium}


We report here the contraction patterns observed in specimen confined in ring-shape chambers. The experiments were realized with 5 different rings of varying width and perimeter. Experiments with every ring has been performed with 16 different specimen and on time scale ranging from 4 to 12 hours. The ring perimeter $L$ (measured at the center) ranges from $6.0$ to $13.5$ cm (see Fig. \ref{fig:Set-up_image}). The aspect ratio $L/e$, with $e$ the width of the ring, ranges from $11$ to $41$, allowing us to consider one-dimensional (1D) description.
In this configuration the plasmodium uniformly fills the gap. Transmitted light imaging allows us to record spatio-temporal thickness variations. 
In order to have tractable datasets, the signal is averaged over angular sectors with arc length $\simeq 1.1~\rev{\text{mm}}$ (\textit{Materials and Methods}).
The signal is then a function of the angular position $\theta$ and time $t$. The height at a given position along the ring is decomposed as $h(\theta,t)=D(\theta, t)+A(\theta,t)\cos \phi(\theta,t)$, where the drift $D$, amplitude $A$ and phase $\phi$ are functions that vary on long, intermediate, and short timescales, respectively (\textit{Materials and Methods}). 
\rev{The drift is then subtracted using moving average over a time window of $200~\text{s}$ (see \textit{Materials and Methods}) to focus on the contractile activity. Variations of the relative height $\Delta h=h(\theta,t)-D(\theta,t)$ can then be represented using space-time plots $\left( \theta, t \right)$. Variations of $h$ on longer timescales will be presented in a separate study.}

\subsection{Contractile activity organize as successive transient coherent modes}
The different contractile modes appearing during experiments are summarized below.
In contrast to what has been reported for protoplasmic droplets \cite{TAKAGI2008420}, we do not observe any specific order in the succession of these modes, which often alternate with chaotic periods of time where contractile activity is not coordinated on the entire ring. \rev{The total time spent in any of these distinct modes is estimated to be $67\%$ of the total time of experiments. The mean duration of every individual mode, as well as its occurrence probability (defined as the total time spent in that mode divided by the total time where coordinated contractions are observed), are summarized in table \ref{table} (see \textit{Materials and Methods} for details on the method to estimate these times).}
%

\begin{table*}[h]
	\centering
	\rev{
\begin{tabularx}{\textwidth}{|Y||Y|Y|Y|Y|Y|}
	\hline
	\textbf{Mode} & travelling & counter-propagating & drifted counter-propagating &  standing & synchronous \\
	\hline
	\hline
	\textbf{Mean duration (sec)} & $1083$ & $2278$ & $6202$ & $1084$ & $1152$ \\
	\hline
	\textbf{Occurrence probability} & $16\%$ & $55\%$ & $21\%$ & $5\%$ & $3\%$ \\
	\hline
\end{tabularx}
}
\caption{\rev{Mean duration and occurrence probability of the different stable modes.}\label{table}}
\end{table*}

\subsubsection*{Travelling wave pattern.}
\begin{figure*}[h]
	\centering
	\begin{subfigure}{1.\columnwidth}
		\includegraphics[width=\textwidth]{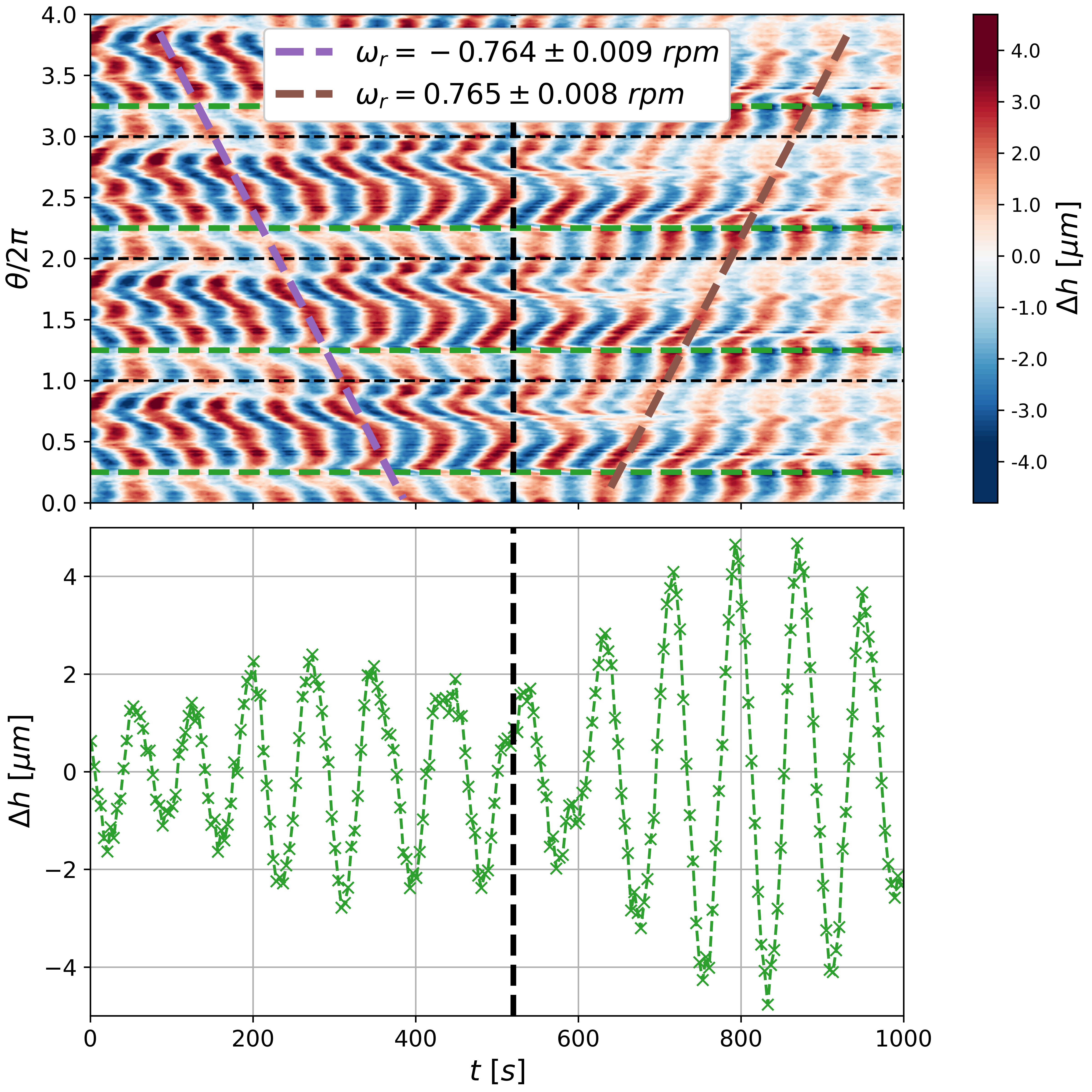}
	\end{subfigure}
	\hfill
	\begin{subfigure}{0.8\columnwidth}
		\includegraphics[width=\textwidth]{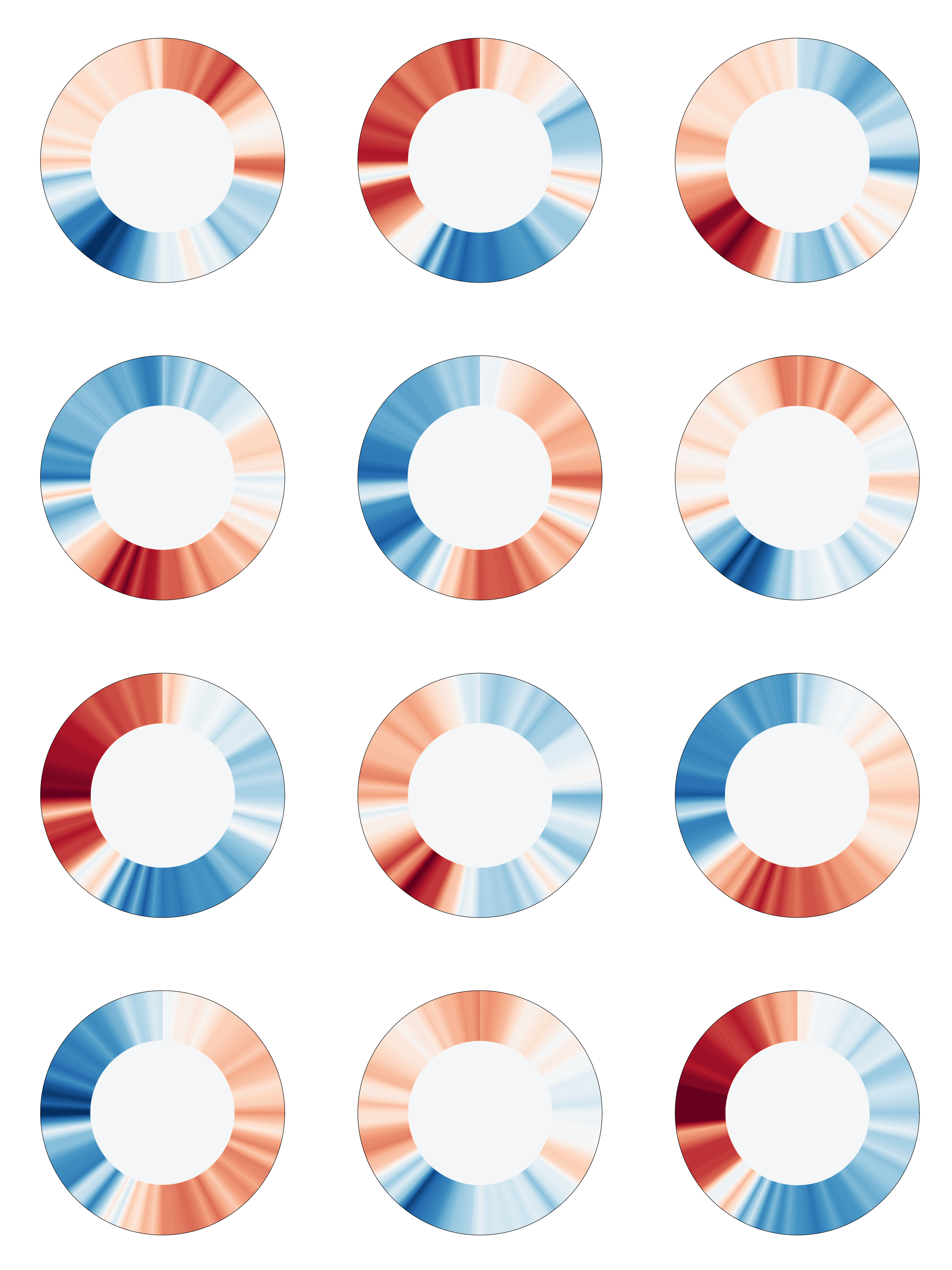}
	\end{subfigure}
	\caption{Top left: space-time plot of travelling wave pattern.  \rev{The spatially periodic signal has been duplicated along the y-axis to help identifying the pattern.} Note the sudden change of wave orientation at $t\simeq 550~\rev{\text{s}}$, highlighted with the vertical line. The wave velocity (expressed in rpm) before and after the change of direction are almost identical. Bottom left: time evolution of relative height $\Delta h$ at a given position ($\theta = \pi /2$, indicated by the horizontal green dashed line on the space-time plot). The fundamental frequency is $f\simeq 0.76~\text{rpm} \simeq 1.2\times10^{-2}~\text{Hz}$. \rev{Right (from left to right, and top to bottom): corresponding representation of height variation along the ring at successive times.} }
	\label{fig:Kymo_traveling}
\end{figure*}

The simplest observed mode, although not the most frequent one, corresponds to a single travelling wave propagating along the ring, in one direction or the other. A typical space-time plot is represented in Fig. \ref{fig:Kymo_traveling}. The contour lines are straight parallel lines and their slope is equal to the wave-velocity. Sudden reversals of the direction of propagation are sometimes observed, as this is illustrated in Fig. \ref{fig:Kymo_traveling}, confirming that there is no rear-front asymmetry in the circular plasmodium. \rev{These reversals occur on a very short timescale (typically, 1 or 2 periods), and several times during an experiment, suggesting that they are not related to a change in network structure.} The frequency of contraction is very similar in both directions ($f\simeq 0.76~\text{min}^{-1}$ for the example shown in Fig. \ref{fig:Kymo_traveling}), and seems independent of the ring dimensions. We will discuss this point hereafter. Note that in a plasmodium free to move, travelling waves are involved in the mechanism  of peristaltic locomotion
\cite{matsumoto_locomotive_2008,rieu_periodic_2015,zhang_self-organized_2017,lewis_analysis_2017}. Here, we see that this mode is generated even in absence of locomotion. However, this mode is rather rare and transient in the full contractile activity observed.  Its duration does not exceed $500~s$, typically $5$ periods of oscillations.

\subsubsection*{Counter-propagating (bidirectional) wave pattern.}
\begin{figure*}[h]
	\centering
	\begin{subfigure}{1.\columnwidth}
		\includegraphics[width=\textwidth]{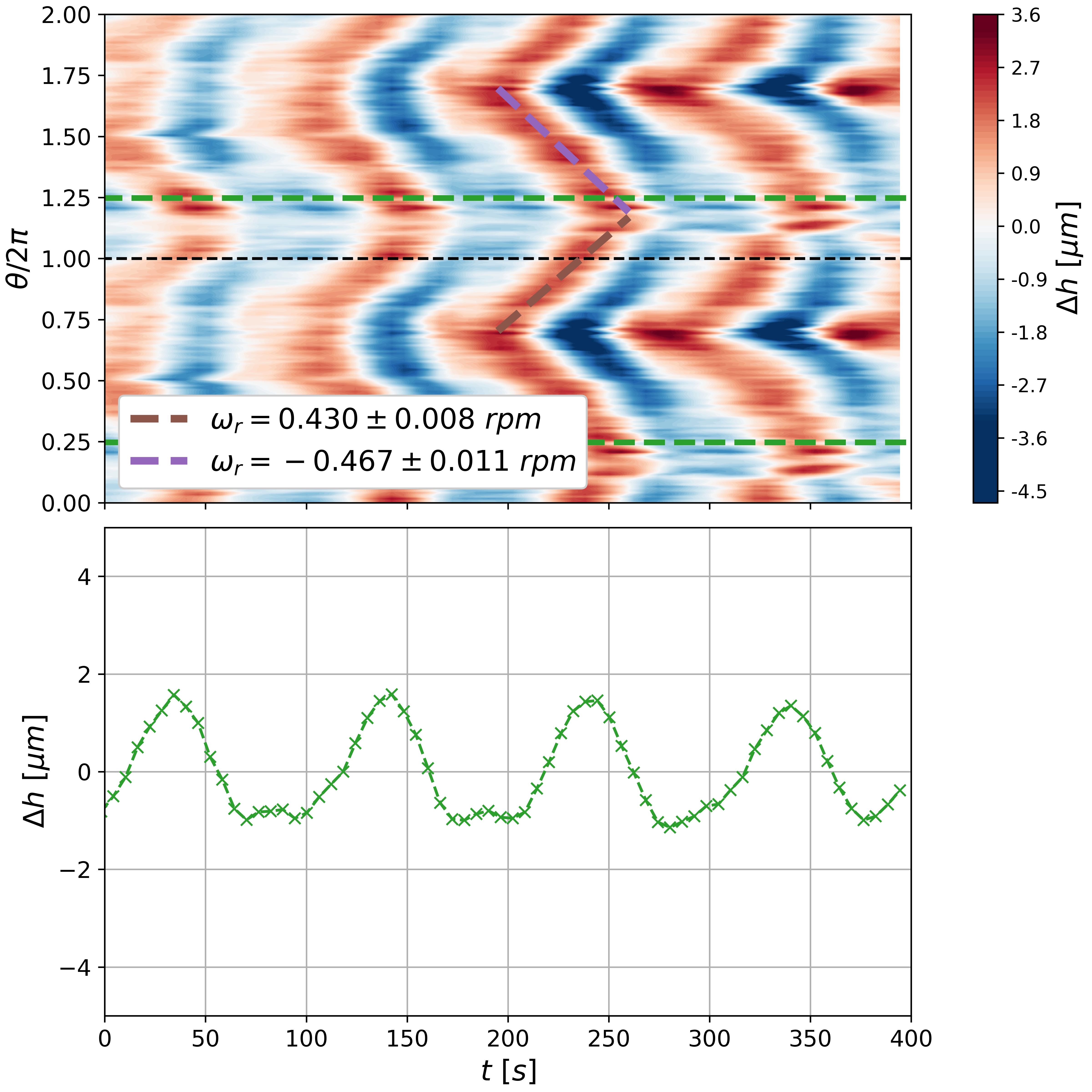}
	\end{subfigure}
	\hfill
	\begin{subfigure}{0.8\columnwidth}
		\includegraphics[width=\textwidth]{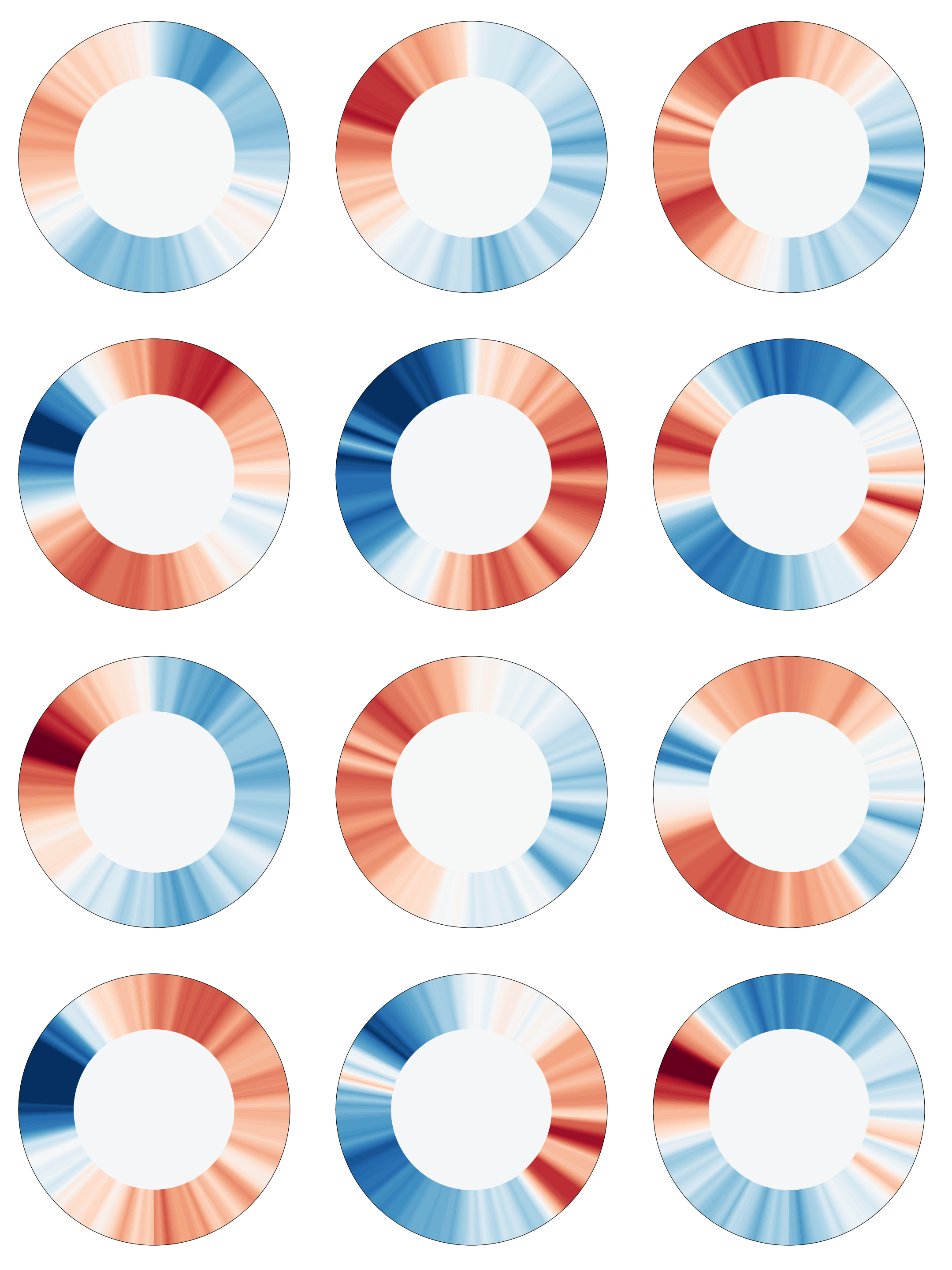}
	\end{subfigure}
	\caption{Top left: space-time plot of counter-propagating wave pattern.  \rev{The spatially periodic signal has been duplicated along the y-axis to help identifying the pattern.} The velocity of the two waves have nearly equal magnitudes, but with opposite sign.  Bottom left: time evolution of the relative height $\Delta h$ at a given position ($\theta = \pi /2$, indicated by the horizontal green dashed line on the space-time plot). The fundamental frequency is $f \simeq 10^{-2}~\rev{\text{Hz}}$. Right (from left to right, and top to bottom): corresponding representation of height variation along the ring at successive times. }
	\label{fig:Kymo_planewave}
\end{figure*}

The most frequently observed mode corresponds in fact to a pair of contraction waves that originate from a given point, travel along the ring at same constant speed but in opposite \rev{directions,} and then annihilate each other at the antipodal point (see Fig. \ref{fig:Kymo_planewave}). 
Successive pairs of counter-propagative waves emerge from the same source point.
%
The space-time plot of this mode exhibits a chevron pattern, as shown in Fig. \ref{fig:Kymo_planewave}. 
At a given position along the ring, the signal oscillates with a typical frequency $f \simeq 1.0\times 10^{-2}~\text{Hz}$, which here again seems independent of the ring dimensions. Interestingly, this pattern was also observed in other 1D excitable media, such as cardiac tissue ring \cite{gonzalez2000resetting,nagai2000paroxysmal} or chick ventricular cells \cite{gonzalez2003reentrant}. 


\subsubsection*{Drifted counter-propagating wave pattern.}
\begin{figure}[h]
	\centering
		\includegraphics[width=\columnwidth]{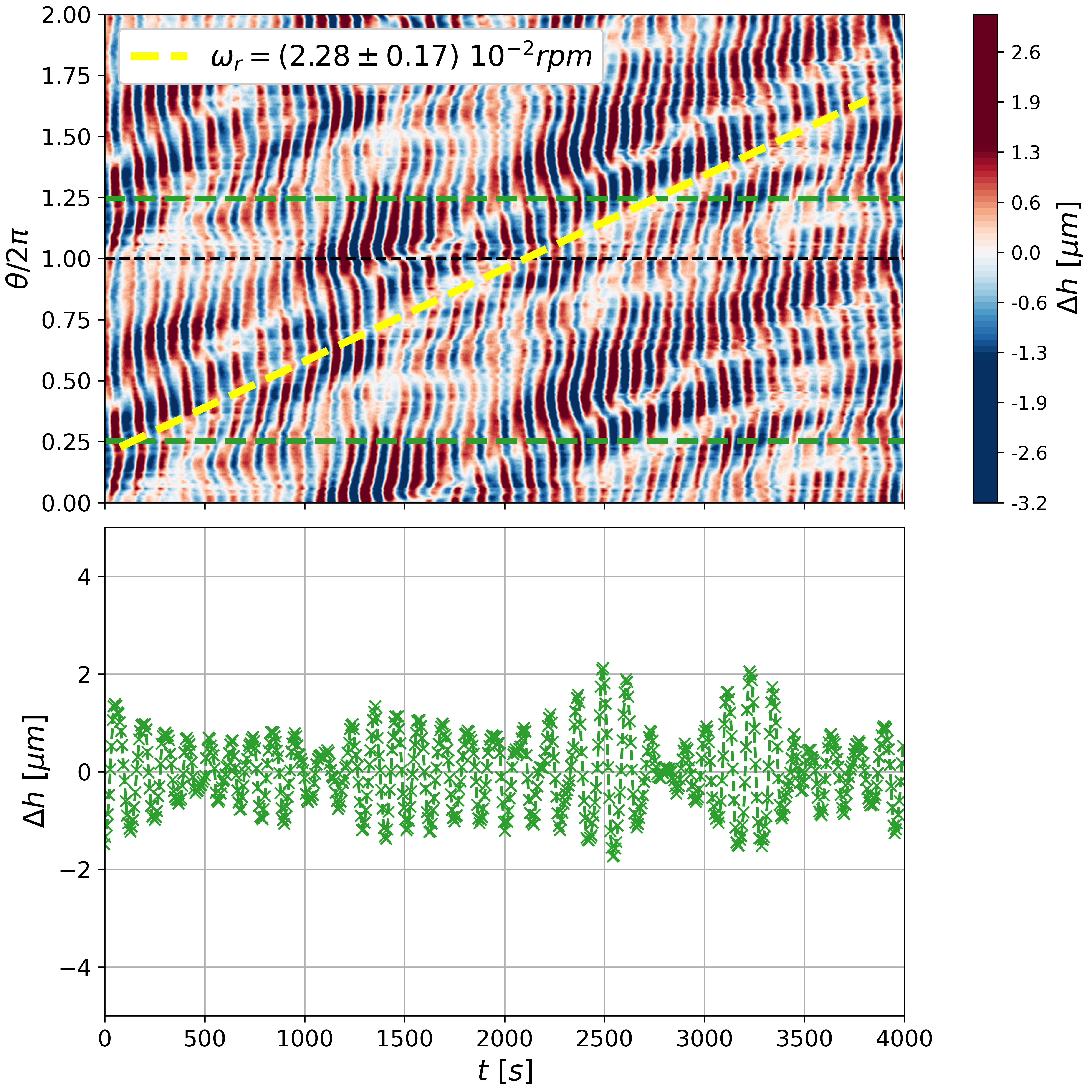}
	\caption{Top : space-time plot of drifted counter-propagating wave pattern.  \rev{The spatially periodic signal has been duplicated along the y-axis to help identifying the pattern.} The angular velocity of the drift (expressed in rad/s) is indicated with a dashed yellow line. Bottom: time evolution of the relative height $\Delta h$ at a given position ($\theta = \pi /2$, indicated by the horizontal green dashed line on the space-time plot). The fundamental frequency is $f \simeq 8.8\times 10^{-3}~\rev{\text{Hz}}$.   \rev{The representation of height variation along the ring, not shown here, is similar to the one obtained for the counter-propagating mode on a few periods (Fig. \ref{fig:Kymo_planewave}).} }
	\label{fig:Kymo_mod_rot}
\end{figure}

We also observe a variant of the counter-propagating wave pattern, in which the generating and annihilating points slowly drift along the ring, in one direction or the other, with sudden change of drift direction (see Fig. \ref{fig:Kymo_mod_rot}). These frequent shifts of drift direction, but also the fact that the drift in a given direction can lasts for several turns, exclude the possibility that the drift is caused by some heterogeneity in the plasmodium composition.

\subsubsection*{Standing wave pattern.}
\begin{figure*}[h]
	\centering
	\begin{subfigure}{1.\columnwidth}
		\includegraphics[width=\textwidth]{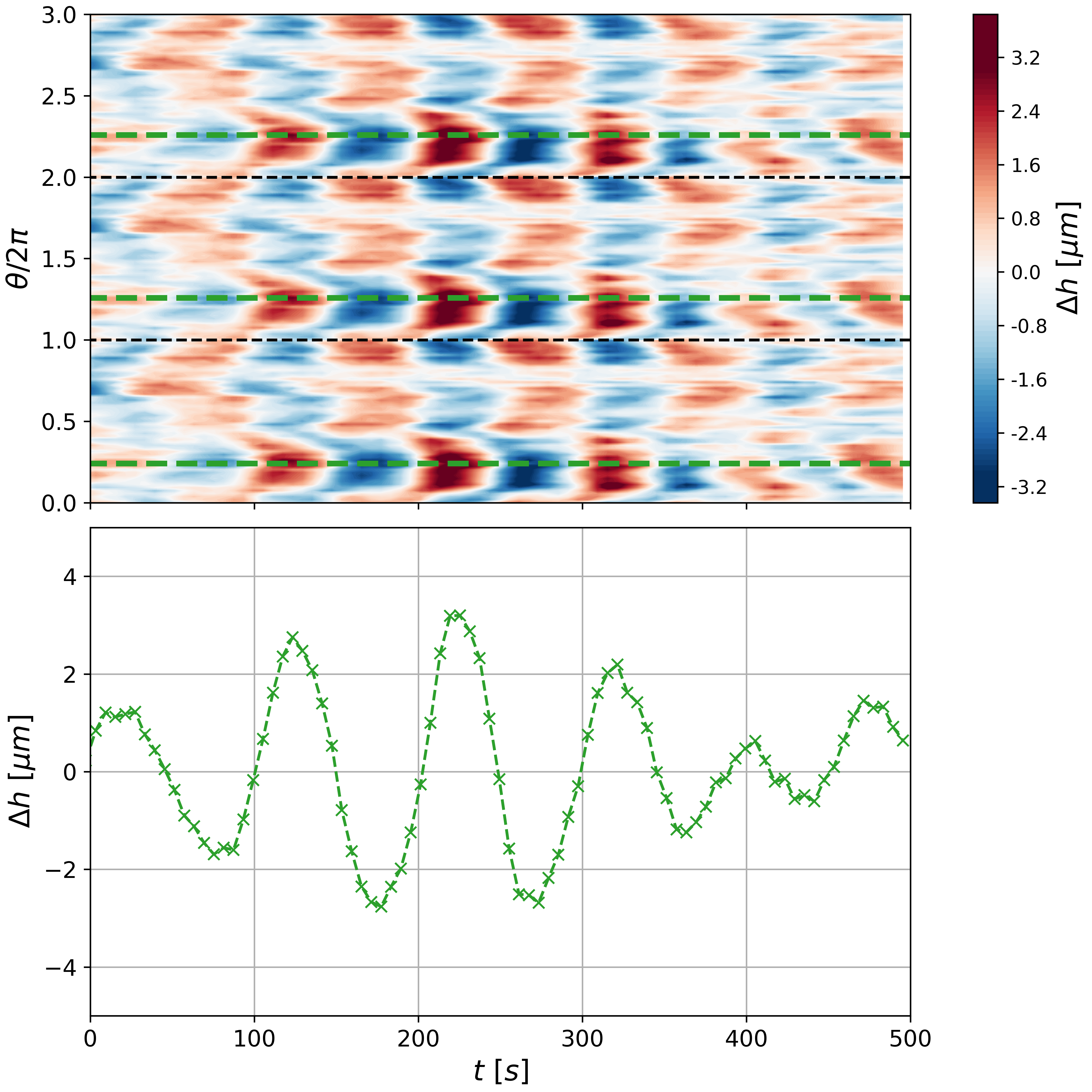}
	\end{subfigure}
	\hfill
	\begin{subfigure}{0.8\columnwidth}
		\includegraphics[width=\textwidth]{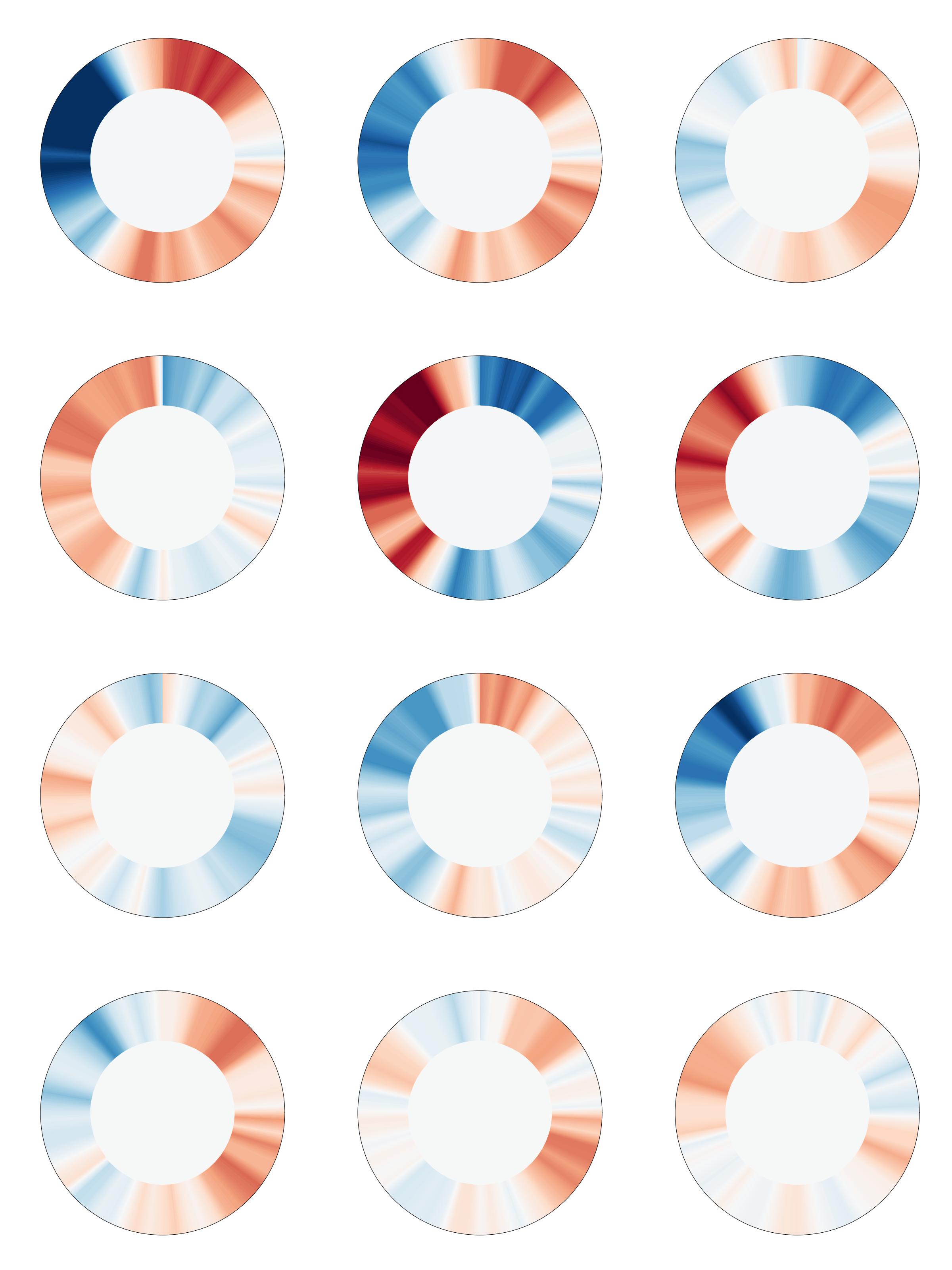}
	\end{subfigure}
	\caption{Top left: space-time plot of standing wave pattern. \rev{The spatially periodic signal has been duplicated along the y-axis to help identifying the pattern.}  Bottom left: time evolution of relative height $\Delta h$ at a given position ($\theta = \pi /2$, indicated by the horizontal green dashed line on the space-time plot). The fundamental frequency is $f \simeq 1.18\times10^{-2}~\rev{\text{Hz}}$. \rev{Right (from left to right, and top to bottom): corresponding representation of height variation along the ring at successive times.} }
	\label{fig:Kymo_standing}
\end{figure*}

Another frequently observed mode is the {standing wave mode}, in which one half of the annular plasmodium is contracting whereas the second half is dilating. A typical space-time plot is shown in Fig. \ref{fig:Kymo_standing}. This mode is generally associated with amphistaltic locomotion \cite{matsumoto_locomotive_2008,rieu_periodic_2015,zhang_self-organized_2017}, although we see here that locomotion is not mandatory to observe this pattern. It can be noticed than the fundamental frequency of the standing wave is very close to the frequency of the travelling wave pattern reported above ($f\simeq 1.18\times 10^{-2}~\text{Hz}$ in the example shown in Fig. \ref{fig:Kymo_standing}).

%

\subsubsection*{Synchronous pattern.}
\begin{figure*}[h]
	\centering
	\begin{subfigure}{1.\columnwidth}
		\includegraphics[width=\textwidth]{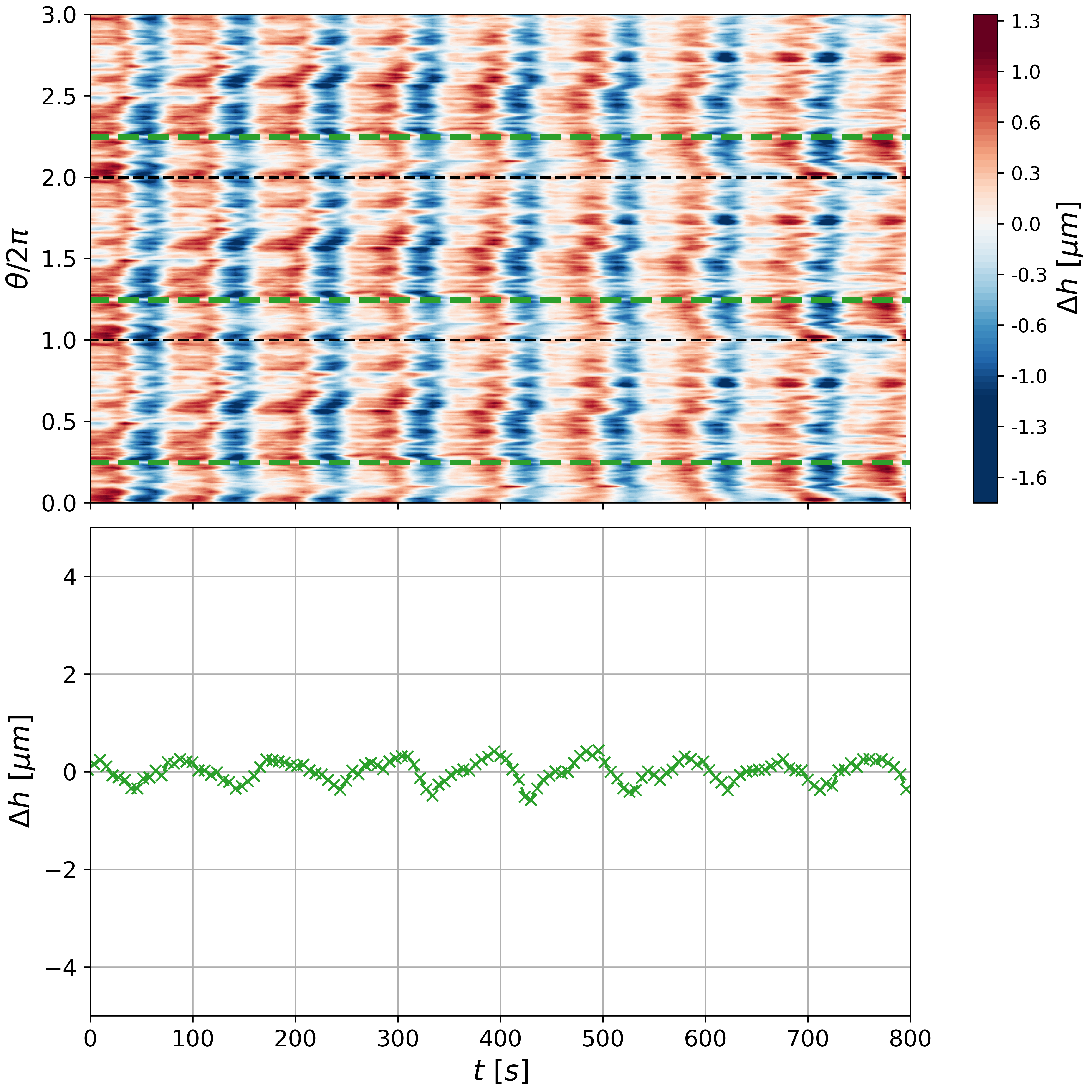}
	\end{subfigure}
	\hfill
	\begin{subfigure}{0.8\columnwidth}
		\includegraphics[width=\textwidth]{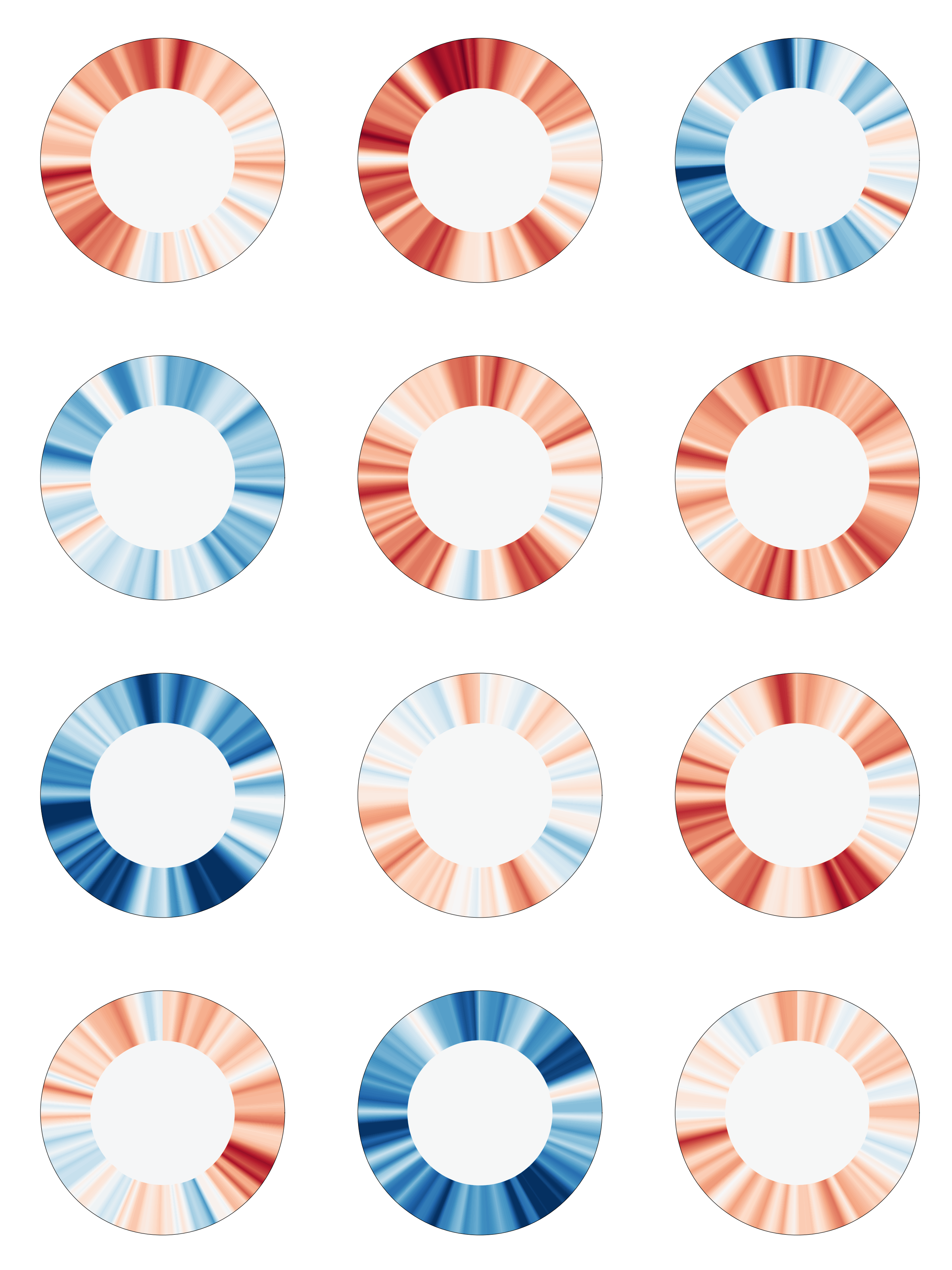}
	\end{subfigure}
	\caption{Top left: space-time plot of synchronous wave pattern. \rev{The spatially periodic signal has been duplicated along the y-axis to help identifying the pattern.} Bottom left: time evolution of relative height $\Delta h$ at a given position ($\theta = \pi /2$, indicated by the horizontal green dashed line on the space-time plot). The fundamental frequency is $f \simeq 0.9\times 10^{-2}~\text{Hz}$. \rev{Right (from left to right, and top to bottom): corresponding representation of height variation along the ring at successive times.} }
	\label{fig:Kymo_sync}
\end{figure*}

Finally, we also observe occasionally a pattern where oscillation seems to take place synchronously in the whole plasmodium ring (figure \ref{fig:Kymo_sync}). At first thought, such a pattern should not be possible as it seems to violate volume conservation.  However, this can be an artefact caused by the signal averaging over the ring width: plasmodium at the ring edges can be in opposite contraction phase with plasmodium in its center, such that the overall volume is conserved. The pattern observed then just reflects the change of concavity of the plasmodium height profile. This would also explain the small amplitude of signal observed for this pattern. Note that such a synchronous pattern with antiphase contraction at the peripheral has also been reported in protoplasm droplets \cite{TAKAGI2008420}.
%

\subsection{Influence of ring geometry on the frequency and wavelength}
The modes described previously were observed in experiments performed with five rings of different size. We analyze here how the ring size affects the contraction patterns. Because the amplitude and frequency of the signal are unsteady, Fourier analysis is inadequate to determine the frequency and wavelength of the wave patterns. \rev{We then use a more adapted signal processing technique to extract the instantaneous phase $\phi(\theta,t)$, initially developed for processing electroencephalography signals (\textit{Materials and Methods}).} From this instantaneous phase we straightforwardly determine the frequency $f$ and the phase velocity $v_{\phi}$, defined respectively as
\begin{equation}
	f =  \partial_t \phi/2\pi,
	\quad \quad \quad
	v_\phi=-r\partial_t \phi /  \partial_\theta \phi,
\end{equation}
where $r=L/2\pi$ is the ring radius. We then define the wavelength of the pattern as $\lambda=v_\phi f$.

\subsubsection*{Frequency is independent of ring size.}
Figure  \ref{fig:wavelength_meanfreq}(a) shows the most probable value of the frequency measured over all data collected as a function of the ring diameter $d=2r$.  The frequency is clearly independent of the ring size, with mean value $f = \left(1.04 \pm 0.18\right)\times 10^{-2}~\text{Hz}$.
Average values of the measured frequency, rather than most probable ones,  show very similar results (not shown).
Note that our value of $f$ is significantly lower than the frequency reported for protoplasmic droplets \cite{TAKAGI2008420}, but in very good agreement with what has been reported elsewhere \rev{\cite{wohlfarth-bottermann_oscillating,PhysRevLett.85.2026,alim_random_2013}.} Influence of some external parameters such as temperature could explain this difference \cite{wohlfarth-bottermann_oscillating}.

\subsubsection*{Wavelength scales linearly with ring size.}
\begin{figure}[h]
	\includegraphics[width=\columnwidth]{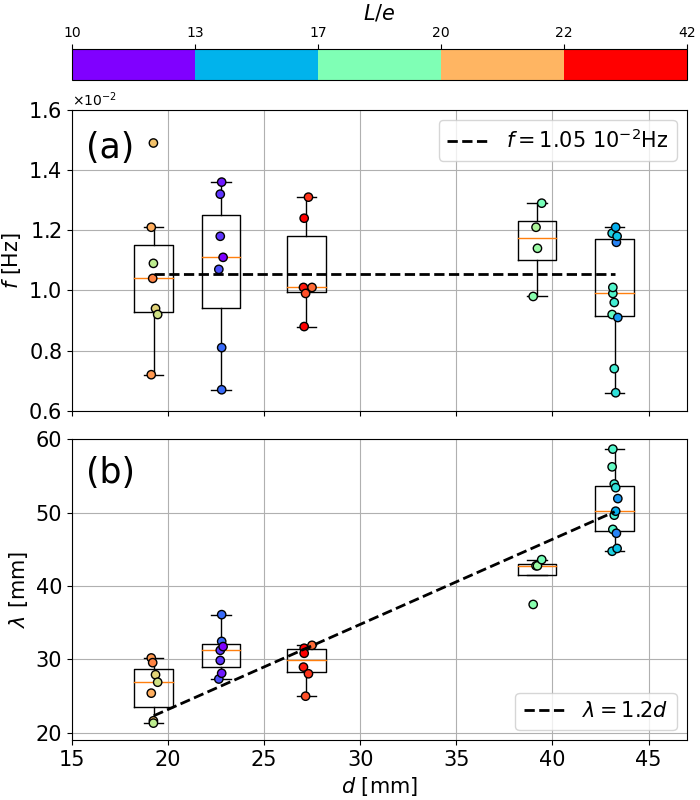}
	\caption{(a) most probable frequency $f$ and (b) most probable wavelength $ \lambda$ as a function of the ring diameter $d=L/\pi$. \rev{Data corresponding to different ring aspect ratios $L/e$ are represented with different colors.} The dashed black line represents the linear fit of the data: $\lambda=1.2d$. Error bars correspond to 3 standard deviations.}
	\label{fig:wavelength_meanfreq}
\end{figure}

Figure \ref{fig:wavelength_meanfreq}(b) shows the most probable value of the pattern wavelength measured over all data collected as a function of the ring diameter. It must be noted that these data correspond mainly to the counter-propagating wave mode. The wavelength $\lambda$ increases linearly with the ring diameter over the whole investigated range, with a slope very close to $1$. This result contradicts some of the models developed for the self-organization of contractile activity, as we will discuss below.
Note that we also used different ring widths $e$, ranging from 2.1 to 11.4 mm, but still $e \ll d$. We did not observe any significant impact on the measured wavelength on this narrow range of width value.

\section{Discussion}

\subsection*{Reported modes are ubiquitous}
Although our confined plasmodia have heterogeneous structures containing both gel and sol phases, typical of mature plasmodia, all the patterns we observe are also observed in homogeneous protoplasmic droplets \cite{TAKAGI2008420}. The only exception is the drifted counter-propagating wave pattern, which has never been reported before, to our knowledge.
Note however that additional patterns, such as spiral patterns, have been identified in free protoplasmic droplets, which are two-dimensional contractile systems, and as such have a richer phenomenology of contractile patterns.
We also note that the travelling wave and standing wave modes, which play fundamental roles for the peristaltic and amphistaltic locomotions  \cite{matsumoto_locomotive_2008,rieu_periodic_2015,zhang_self-organized_2017,lewis_analysis_2017}, respectively, are also observed in our plasmodium in which growth and displacement are stopped. This suggests that the flow boundary conditions at the plasmodium peripheral are crucial to explain the generation (or not) of displacement.

\rev{Whereas we have clearly identified the fundamental standing wave mode only, having two nodes, multinodal standing wave patterns have also been reported in tadpole-shaped migrating plasmodia \cite{rodiek_patterns_2015}. Such modes, if they ever exist in our case, have very short lifetime and thus cannot be discriminated from chaotic periods.	
 }  

\rev{More generally, the different observed modes are often associated with some biological function, such as locomotion, morphological adaptation, or signal spreading inside the plasmodium \cite{rodiek_patterns_2015,fleig_emergence_2022}, and hence are selected and/or consolidated by specific triggering factor (e.g.: nutrients, chemicals or light). In that perspective, a resent study \cite{fleig_emergence_2022} analyses the chaotic periods that we observed between well-identified patterns as continuous distributions of modes from which any specific behavioural dynamics can emerge.}

\rev{It is also worth noticing that the linear scaling we observed between wavelength and plasmodium size has already been reported and rationalized for extended plasmodia \cite{alim_random_2013}: a contraction pattern with phase growing linearly from 0 to $2\pi$ over the organism length optimizes transport. Note however that for migrating plasmodia Kuroda \etal \cite{kuroda_allometry_2015} found a logarithmic scaling instead.} 


\subsection*{Experimental results support mechano-bio-chemical feedback with self-sustained calcium oscillations}
Our experimental observation of the different patterns can help to discriminate between the theoretical models which have been developed to explain the spatiotemporal self-organization of contractile activity.
Although all models agree on the essential role of calcium Ca$^{2+}$ in the regulation of contraction of the actomyosin cortex in \textit{Physarum polycephalum}, they diverge on the exact coupling mechanisms between contraction and calcium concentration.
They can be divided in two main categories: those relying on a mechano-chemical feedback: 
calcium acts on the contraction, which in turn modifies calcium concentration via mechanosensitive channels \rev{triggered by the stretching of the membrane in which they are embedded} \cite{teplov_continuum_nodate,teplov_role_2017,radszuweit_intracellular_2013,julien_oscillatory_2018}.
Diffusion and advection of calcium also modify its local concentration. In these models, the calcium concentration does not oscillate in absence of contractile activity.
Other models \cite{kobayashi_mathematical_2006,radszuweit_model_2010,radszuweit_active_2014} rely on a mechano-bio-chemical coupling: calcium concentration is regulated by biological process \cite{smith_model_1992} that results in autonomous calcium oscillations, which in turn will act on the contraction of the cortex. Advection and diffusion of calcium eventually lead to spatial coordination of the contractions. In those models, autonomous calcium oscillations occur even in
the absence of a mechanical feedback. 
A few studies have made an exhaustive numerical research of the stable contraction pattern that could emerge from their model in simple geometries. In the first category of models, Julien \& Alim \rev{ \cite{julien_oscillatory_2018}} made an exhaustive numerical study of the patterns predicted by the 1D version of their model, either with periodic or fixed boundary conditions. In the second category, Radszuweit \etal \cite{radszuweit_active_2014} solved numerically a 2D version of their model.

Both models predict some of the patterns that we observe in our experiments:  this is the case of the travelling wave pattern as well as the counter-propagating wave pattern. Note that radial waves in 2D geometries like protoplasmic droplets \cite{radszuweit_active_2014} are the equivalent of the counter-propagating waves we observe in our 1D geometry. Moreover, counter-propagating waves is the most prominent mode in models as in experiments.
However, in our experiments the wavelength increases with the ring perimeter (at least up to a perimeter equal $13.5~\rev{\text{cm}}$), while in the mechano-chemical model \cite{julien_oscillatory_2018} with periodic boundary conditions, the ``wave size'' reaches an asymptotic value independent of the channel channel length when $L$ is larger than the most unstable wavelength $\lambda_{lin} = 7.1~\rev{\text{mm}}$, \rev{except for the rare travelling wave mode, as Figs. 2 and S6 of  \cite{julien_oscillatory_2018} suggest}. In the mechano-bio-chemical model  \rev{\cite{radszuweit_active_2014}} on the other hand, the wavelength is of the same order of magnitude of the protoplasmic droplet, although a systematic analysis with varying size has not been done.



Furthermore, the other patterns -- standing wave pattern and synchronous pattern -- are also expected by the Radszuweit \etal model, whereas they have not been observed within the Julien \& Alim model. Only the drifted counter-propagating wave pattern is not predicted by any model. 

All these results are in favor of models based on the existence of self-sustaining calcium oscillations. Actually, a whole range of experimental evidence in favor of a primary calcium oscillator prior to mechanical contractions have been reported in literature:
\textit{i)} calcium oscillations are observed in situations in which the mechanical oscillations are stopped \rev{\cite{yoshimoto_ca2_1982,miyake_entrainment_1992};}
\textit{ii)} contraction waves propagate synchronously in the vein networks and the rest of the plasmodium, while they have very contrasted permeability to flow \cite{Busson2022};
\textit{iii)} calcium oscillates in antiphase with contraction \cite{yoshimoto_simultaneous_1981}, \rev{whereas mechano-chemical models predict that these quantities oscillate in phase \cite{teplov_continuum_nodate,julien_oscillatory_2018} };
\textit{iv)} when the plasmodium is cut into small pieces, each soon starts oscillating with the same frequency as before \cite{tero_coupled-oscillator_2005}.

Note that a recent extension of the Radszuweit \textit{et al.} model incorporated both feedback mechanisms \cite{alonso_oscillations_2016}. The authors report the emergence of stable contraction patterns with any of the two mechanisms, but self-sustained calcium oscillations enhances the region of pattern formation in comparison to a model that neglects the contribution of such biochemical oscillator.
In order to elucidate completely the feedback mechanism at the origin of these patterns, future research will focus at imaging simultaneously contractile activity, cytoplasmic flow, and calcium waves. 
Besides, the modulation of the amplitude function $A(\theta,t)$ itself shows interesting patterns that also deserves further investigation.

\section{Materials and Methods}
\subsection{Experimental system culture and insemination}
The work presented concerns \emph{Physarum~polycephalum} in the plasmodial phase of its life cycle. In this stage, the specimen morphology is essentially a 2D growing and exploring network or an homogeneous tissue from which the network emerges. Sclerotia were obtained from Carolina Biological, South Carolina, USA. Specimens are essentially grown on aqueous gels made with $2\%$ phytagel, $1\%$ glucose and $10^{-5}$ diluted $CaCl_{2}$ solution in MilliQ Water. They are feeded using sterilized oat flakes disseminated on the gel. 
The Petri dishes are closed and sealed with Parafilm to keep the gel from drying.
The dishes are maintained in a dark chamber with a $25^{\circ} \rev{\text{C}}$ controlled temperature.
Twice a week the growing front of the protoplasm is transferred to a new feeding gel in order to maintain the organism in a constant feeding stage.
%
%
To make ring shape specimen, we directly crop ring shape patches from homogeneous sample using plastic homemade punches. These chambers are initially half filled with $2 \%$ phytagel gel, and thermalized at $25^{\circ} \rev{\text{C}}$ during all experiment. Video recording is started after a settling time of 15 to 30 $\rev{\text{min}}$.

\subsection{Image acquisition}

Images of ring confined \emph{Physarum~polycephalum} are obtained using transmitted light imaging using a Leica Z16 APO macroscope. The lighting is assured by a TL5000 LED lightning base that provides uniform, dimmable and stable lightening.
Acquisition is made using a CMOS Basler color camera (acA2440-75uc) with an exposure time adjusted to minimize light intensity on the sample.
Images are acquired using either 4 or 6 seconds per frame, with a spatial resolution of $37~\rev{\text{\mu m / pix}}$ for a total film duration of 4 to 12 hours and a resolution of $2448\times 2048~\rev{\text{pix}}^2$.

\subsection{Beer-Lambert law}
To measure the absorbance of \textit{Physarum polycephalum}, we built a wedge with two glass plates in which we placed a large quantity of mixed plasmodium, yielding a controlled and homogeneous thickness gradient. A measure of the transmitted light intensity $I$ along the wedge straightforwardly provides the relation between intensity $I$ and thickness $h$. The data are then fitted using the Beer-Lambert law $I=I_0 e^{-h/\ell_a}$, 
allowing to extract the absorption length $\ell_a$.  We analyze each color channel separately, yielding three different values:  $\ell_a\simeq 100~\rev{\text{\mu m}}$ (blue), $200~\rev{\text{\mu m}}$ (green), $350~\rev{\text{\mu m}} (red)$. 
Although the exact values change with aging of Physarum pigments, they are consistent with the value $\ell_a\simeq 67~\rev{\text{\mu m}}$ reported by Bykov et al. \cite{Bykov2009} of  using Doppler optical coherence tomography technique with an infrared  ($840~\rev{\text{nm}}$) superluminescent diode.

\subsection{Signal analysis}
The height signal is derived from intensity signal using the Beer-Lambert law using the blue channel of the camera, for which \phyp has highest absorbance, and so offers higher sensitivity to thickness variation. 
The height signal is then smoothed by convoluting it with a top hat window bounded by fast Gaussian rise and fall with a total size of $10~rev{\text{frames}}$ (around $60~\rev{\text{s}}$). This filters out every small time scale features, typically those below one half of the period.

Because the signal is unsteady and pseudo-periodic, Fourier analysis is not adapted to capture the instantaneous phase of the signal.
Instead, the following decomposition of the height
at a given position is considered:
\begin{equation}
	h(t) =D(t) + A(t) \cos \phi(t).
\end{equation}
Actually, this decomposition is not unique but the physical meaning of these variables constrains the time scales on which they vary:
\begin{itemize}
	\item The drift $D(t)$ around which the values oscillate corresponds to the slow variation of the baseline diameter of the veins (or the baseline thickness of the cytoplasm).
	\item The amplitude $A(t)$ of the oscillations which modulates the shape of the oscillating signal.
	It varies on intermediate time scales, and may be the vector of chemical information transport \cite{alim_mechanism_2017}.
	\item The instantaneous phase $\phi(t)$ which locates the oscillation in a cycle. It is a fast variable that characterizes the contractile state.
\end{itemize}


\rev{The drift $D(t)$ is obtained using a moving 
	average over a time window of $200~\text{s}$. The phase $\phi(t)$ and amplitude $A(t)$ are then derived using a technique initially developed for quantitative electroencephalography  \cite{etevenon1999high} and adapted to \textit{P. polycephalum} contraction activity \cite{alim_mechanism_2017,bauerle_spatial_2017}:  $\phi (t)$ and $A(t)$ are calculated respectively as the argument and the norm of a complex analytic signal which is built from the Hilbert transform of the relative height variation $h(t)-D(t)$.}
The three variables are then averaged over all the pixels that belong to a same angular sector of the ring-shape chamber. The number $N$ of sectors is adapted to the ring perimeter such that $L/N$ is constant ($\simeq 1.1~\rev{\text{mm}}$).


\rev{
\subsection{Estimation of pattern duration and occurrence}
}
\rev{
The identification of the contractile patterns and the estimation of their duration and occurrence probability are based on the following criteria:
\begin{itemize}
\item We consider that the travelling mode corresponds to patterns for which we are able to follow a phase circulation around the ring over several periods.
\item Standing and contra-propagative modes are sometimes difficult to distinguish. We systematically disregard from the latter patterns having two nodes but no propagation pattern from one node to another. 
\item Drifted contra-propagative mode corresponds to a contra-propagative mode with a slow drift on the source point. This slow drift makes it particularly easy to identify.
\item We systematically disregard from synchronous mode patterns in which a source or node point can be detected.
\end{itemize}
With all these criteria, a stable coordinated contraction pattern developed on the entire ring is observed over $67\%$ of the total recording time, and the estimated duration and occurrence probability of every mode are reported in Table \ref{table}.
}


\ack
We acknowledge financial support from French National Research Agency Grants ANR-17-CE02-0019-01-SMARTCELL and CNRS MITI ‘Mission pour les initiatives transverses et interdisciplinaires’ (reference: BioRes).

\section*{References}
\bibliographystyle{iopart-num}
\providecommand{\newblock}{}

\end{document}